\documentclass[a4paper,fleqn,usenatbib]{mnras}

\usepackage[T1]{fontenc}
\usepackage{ae,aecompl}

\usepackage[normalem]{ulem}
\usepackage{amsmath}
\usepackage{graphicx} 
\usepackage{lscape}
\usepackage{indentfirst}
\usepackage{enumitem}
\usepackage{xspace}

\usepackage{amssymb}
\usepackage[dvipsnames]{xcolor}
\usepackage{url}
\usepackage[flushleft]{threeparttable}

\newcommand {\bc}{\begin {center}}
\newcommand {\ec}{\end {center}}
\newcommand {\be}{\begin {equation}}
\newcommand {\ee}{\end {equation}}
\newcommand {\beq}{\begin {eqnarray}}
\newcommand {\eeq}{\end {eqnarray}}
\newcommand {\comment}[1]{}

\newcommand {\ergs}{{\rm erg\ \rm s^{-1}}}

\newcommand{\swift}{Swift~J0243.6+6124 }
\newcommand{\RX}{RX~J0209.6-7427 }

\title[X-ray pulsars in the neutrino sky]
{
Bright X-ray pulsars as sources of {MeV} neutrinos {in the sky}
}
\author[A.~Asthana et al.] 
{Aman~Asthana$^{1}$\thanks{E-mail: amanasthana12345@gmail.com (AA)},
Alexander A. Mushtukov$^{2}$\thanks{E-mail: alexander.mushtukov@physics.ox.ac.uk (AAM)},
Alexandra A. Dobrynina$^{3}$
and
Igor S. Ognev$^{3}$
\\ 
$^1$ Coppell High School, 185 W Parkway Blvd, Coppell, TX 75019 Coppell, USA \\ 
$^2$ Astrophysics, Department of Physics, University of Oxford, Denys Wilkinson Building, Keble Road, Oxford OX1 3RH, UK\\
$^3$ P.G. Demidov Yaroslavl State University, Sovietskaya 14, 150003 Yaroslavl, Russia \\
} 

\pubyear{2023}

\begin{document}
\label{firstpage}
\pagerange{\pageref{firstpage}--\pageref{lastpage}}
\maketitle


\begin{abstract} 
{High mass accretion rate onto strongly magnetised neutron stars results in the appearance of accretion columns supported by the radiation pressure and confined by the strong magnetic field of a star. 
At mass accretion rates above $\sim 10^{19}\,{\rm g\,s^{-1}}$, accretion columns are expected to be advective.
Under such conditions, a noticeable part of the total energy release can be carried away by neutrinos of a MeV energy range.
Relying on a simple model of the neutrino luminosity of accreting strongly magnetised neutron stars, we estimate the neutrino energy fluxes expected from six ULX pulsars known up to date and three brightest Be X-ray transits hosting magnetised neutron stars.
Despite the large neutrino luminosity expected in ULX pulsars, the neutrino energy flux from the Be X-ray transients of our Galaxy, SMC and LMC is dominant.
However, the neutrino flux from the brightest X-ray transients is estimated to be below the isotropic background by two orders of magnitude at least, which makes impossible direct registration of neutrino emission from accreting strongly magnetised neutron stars nowadays.}
\end{abstract}


\section{Introduction}
\label{sec:Intro}

X-ray pulsars (XRPs) are strongly magnetised neutron stars (NSs) in close binary systems (see, e.g., \citealt{2022arXiv220414185M}).
Their emission in X-rays is caused by the accretion of matter from a companion star, which loses its mass via stellar wind or through the inner Lagrange point.
A strong magnetic field in XRPs affects the geometry of accretion flow directing it towards small regions located close to magnetic poles of a NS and details of the interaction between material and radiation \citep{2006RPPh...69.2631H}. 
The apparent luminosity of XRPs covers orders of magnitude from $10^{32}\,\ergs$ to $10^{41}\,\ergs$.
The brightest XRPs belong to the recently discovered class of pulsating ultra-luminous X-ray sources (ULXs, see \citealt{2014Natur.514..202B,2017Sci...355..817I} and \citealt{2021AstBu..76....6F} for review). 
The relation between the actual and apparent luminosity in the brightest ULXs is still under debate.
On one hand, extreme accretion onto a NS can cause strong outflows from accretion disc \citep{1973A&A....24..337S}, which naturally results in geometrical collimation of X-ray radiation \citep{2007MNRAS.377.1187P,2009MNRAS.393L..41K,2017MNRAS.468L..59K}.
According to some authors, the collimation of X-ray radiation can lead to the fact that in certain directions, the apparent luminosity of the ULXs has orders of magnitude higher than the actual one \citep{2017MNRAS.468L..59K,2020MNRAS.494.3611K}.
On the other hand, the high pulsed fractions observed in pulsating ULXs largely excludes strong geometrical beaming \citep{2021MNRAS.501.2424M,2023MNRAS.518.5457M}, which says in favour of a relatively small difference between the luminosities. 
The possible similarity between the actual and apparent luminosity in ULX hosting NSs is in agreement with the recent result of population synthesis models \citep{2020AstL...46..658K} and the analysis based on the observed luminosity function of high mass X-ray binaries \citep{2015MNRAS.454.2539M}.

High mass accretion rates ($\gtrsim 10^{17}\,{\rm g\,s^{-1}}$) onto the surface of magnetised NSs results in the appearance of radiation pressure dominated shock above the surface and accretion columns \citep{1976MNRAS.175..395B,1981A&A....93..255W,2015MNRAS.447.1847M,2015MNRAS.454.2539M} - structures supported by radiation pressure and confined by a strong magnetic field.

Under conditions of very high mass accretion rates ($\gtrsim 10^{19}\,{\rm g\,s^{-1}}$), accretion columns turn into advection dominated state when the typical time of photon diffusion from the flow becomes comparable or larger than the dynamical timescale of accretion process \citep{2018MNRAS.476.2867M}.
In this case, a large fraction of energy can turn into the production of electron-positron pairs 
\citep{2019MNRAS.485L.131M} and further neutrino emission due to the pair's annihilation \citep{1992PhRvD..46.4133K}.
As a result, the brightest XRPs can be comparably bright in neutrinos.

The theory of neutrino emission from bright XRPs requires the development of an accurate model of accretion column, which accounts for mechanisms of magnetic opacity \citep{2022MNRAS.517.4022S,2023IAUS..363..327S} and possibly deviations of accreting matter state from the thermodynamic equilibrium (see Discussion in \citealt{2019MNRAS.485L.131M}). 
Nonetheless, some order of magnitude estimations can be done in the base of the toy model of the accretion column with advection \citep{2018MNRAS.476.2867M}.
The most promising candidates for the role of neutrino pulsars are pulsating ULXs (assuming their apparent luminosity is close to the actual one) and bright Be X-ray transients, which reach luminosity $\sim 10^{39}\,\ergs$ at the peak of their outbursts  (see \citealt{2011Ap&SS.332....1R} for review). 

In this paper, we consider six pulsating ULXs known up to date -
M82~X-2 \citep{2014Natur.514..202B}, 
NGC~5907~X-1 \citep{2017Sci...355..817I},
NGC~7793~P13 \citep{2016ApJ...831L..14F,2017MNRAS.466L..48I},
NGC~300~X-1 \citep{2018MNRAS.476L..45C},
NGC~1313~X-2 \citep{2019MNRAS.488L..35S}, and
M51~X-7 \citep{2020ApJ...895...60R}
- and three of the brightest Be X-ray transients:
\swift \citep{2020MNRAS.491.1857D},
SMC~X-3 \citep{2017A&A...605A..39T},
and 
\RX \citep{2020MNRAS.494.5350V,2022ApJ...938..149H}.
Using a toy model of neutrino emission from accreting strongly magnetised NSs proposed by \citealt{2018MNRAS.476.2867M}, we estimate the neutrino luminosity of considered objects and their corresponding neutrino energy flux at Earth. 
We compare the estimated flux level with the flux of neutrino background and sensitivity of neutrino telescopes. 

\begin{table*}
\centering
\caption{
The table represents the maximal apparent X-ray luminosity of the objects $L_{X}$, 
expected neutrino luminosity $L_\nu$, 
distance to the source, 
and corresponding neutrino energy flux $F_\nu$ at Earth from six pulsating ULXs discovered up to date and three bright Be XRPs.
Estimations of neutrino luminosity are performed under the assumption that the apparent X-ray luminosity is equal to the actual one.
NS mass and radius are taken to be $M=1.4M_\odot$ and $10^6\,{\rm cm}$ respectively. 
}
\tabcolsep=0.3cm
  \label{tab:ULXs}
  \begin{tabular}{l r r c c c c}
    \hline\hline
    \vspace{0.1cm}
  Name & $L_{X}$ (max) & $L_\nu$  & $D$ & $F_\nu$  \\
       & $[\ergs]$  & $[\ergs]$ & $[{\rm ly}]$ & $[{\rm MeV\,cm^{-2}\,s^{-1}}]$  \\
    \hline 
    NGC~5907~X-1 & $1.5\times10^{41}$ & $3.5\times 10^{41}$ & $5.4\times 10^7$ & $6.7\times 10^{-6}$ \\
    NGC~1313~X-2 & $2\times 10^{40}$ & $1.3\times 10^{40}$ & $1.3\times 10^7$ & $4.3\times 10^{-6}$ \\    
    M82~X-2 & $1.8\times 10^{40}$ & $9\times 10^{39}$ & $1.1\times 10^7$ & $3.9\times 10^{-6}$ \\
    M51~X-7 & $7\times10^{39}$ & $8.4\times 10^{38}$ & $2.5\times 10^8$ & $7.4\times 10^{-10}$  \\
    NGC~7793~P13 & $5\times 10^{39}$ & $4\times 10^{38}$ & $1.2\times 10^7$ & $1.5\times 10^{-7}$ \\
    NGC~300~X-1 & $5\times10^{39}$ & $4\times 10^{38}$ & $6\times 10^6$ & $6\times 10^{-7}$  \\
    \hline
    SMC~X-3 & $2.5\times 10^{39}$ & $9\times 10^{37}$ & $2\times 10^5$ & $1.3\times 10^{-4}$ \\
    \swift & $2\times 10^{39}$ & $5.7\times 10^{37}$ & $1.64\times 10^4$ & $1.2\times 10^{-2}$ \\
    \RX & $1.1\times 10^{39}$ & $1.7\times 10^{37}$ & $2\times 10^5$ & $2.3\times 10^{-5}$ \\
    \hline\hline
  \end{tabular}
\end{table*}

\section{Model}
\label{sec:Model}

\begin{figure}
\centering 
\includegraphics[width=8.9cm]{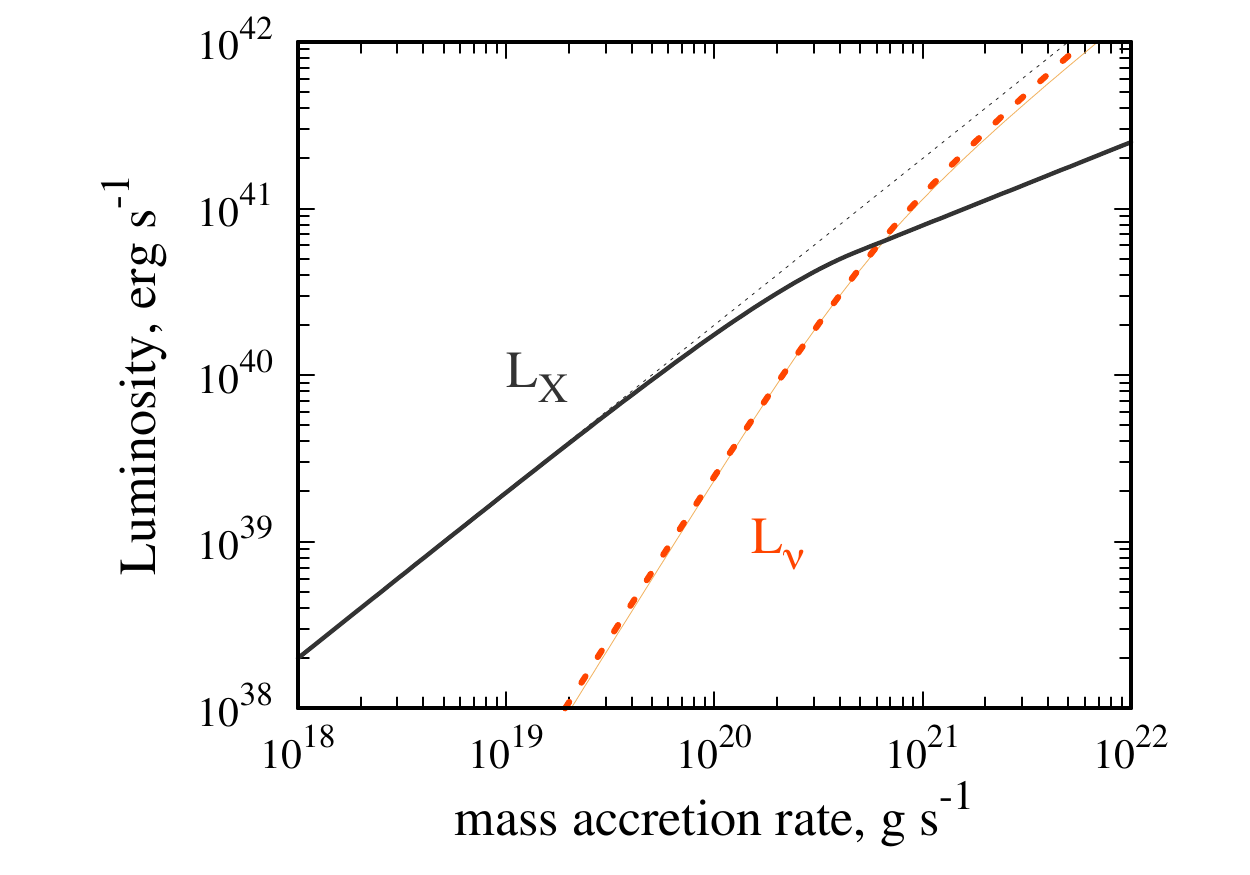} 
\caption{
Dependence of X-ray (black solid line) and neutrino (red dashed line) luminosity on the mass accretion rate onto NS surface \citep{2018MNRAS.476.2867M}.
The solid orange line approximates neutrino luminosity according to (\ref{eq:approx}).
Total luminosity is given by the dotted black line. 
One can see that the total luminosity is dominated by photons at mass accretion rates $\lesssim 2\times 10^{20}\,{\rm g\,s^{-1}}$.
At higher mass accretion rates, neutrino luminosity becomes comparable to the photon luminosity and dominates at $\dot{M}\gtrsim 10^{21}\,{\rm g\,s^{-1}}$.
}
\label{pic:model}
\end{figure}

The estimation of neutrino luminosity and flux is based on the model proposed by \citealt{2018MNRAS.476.2867M}.
The model considers super-critical X-ray pulsars \citep{1976MNRAS.175..395B}, where the luminosity of an accreting NS is high enough to support radiation pressure dominated accretion columns above NS magnetic poles.
If the mass accretion rate onto a NS is high enough, the optical thickness of the accretion column prevents photon excitation from the accretion flow making the flow advective. 
Under these conditions, the internal temperature can increase up to a few hundred keV or even a MeV.
It results in the production of electron-positron pairs and further neutrino emission due to the pair's annihilation
$e^-e^+ \to \nu \, \overline\nu$. 
Neutrinos produced due to the pair annihilation belong to the MeV energy band and are dominated initially (before accounting for neutrino oscillation) by electron neutrinos and anti-neutrinos.
Neutrino synchrotron emission: $e^\mp \to e^\mp \nu \, \overline\nu$ also contributes to the total neutrino luminosity, but its contribution is expected to be relatively small~\citep{1992PhRvD..46.3256K}.
{
The major uncertainty of the model is due to the fact that the model relies on the assumption that the concentration of electron-positron pairs is close to the equilibrium one. 
This condition may not be fulfilled in the accretion column because of short dynamical times of accretion ($\sim 10^{-4}\,{\rm s}$), which may be insufficient for the equilibrium concentration of the pairs (see discussion in Section\,2 of \citealt{2019MNRAS.485L.131M}). 
If the concentration of pairs is lower than the equilibrium concentration, the neutrino emission losses are also lower.
Therefore, the estimates on which we rely should be regarded as optimistic.
Uncertainties associated with the geometry of the accretion channel at the NS surface lead to uncertainties of the neutrino luminosity determination within a magnitude of $2-3$.
}

The expected neutrino luminosity {calculated under the assumption of the equilibrium pair concentration in the accretion channel} can be approximated as 
\beq \label{eq:approx}
L_\nu \approx 0.64\,L_{\rm tot}\,\arctan\left(\frac{L_{\rm tot}}{8\times 10^{40}\,\ergs}\right),
\eeq 
where $L_{\rm tot}=L_{\rm X}+L_\nu$ is the total luminosity of the source due to photon and neutrino emission (see 
Fig.\,\ref{pic:model} and more accurate results in~\citealt{2018MNRAS.476.2867M}).

Both photon and neutrino luminosity experience gravitational redshift.
The luminosity in the observer's reference frame $L_\infty$ is related to the luminosity at the surface of a NS of mass $M$ and radius $R$ as
\beq
\label{eq:red_shift}
L_\infty = L(1-u),
\eeq
where the compactness $u=R_{\rm S}/R$, the Schwarzschild radius of a NS $R_{\rm S}={2GM}/{c^2}\approx 3\times 10^5({M}/{M_\odot})\,{\rm cm}$, and $M_\odot$ is the mass of the Sun.
The dependence of neutrino luminosity $L_\nu$ on the total luminosity $L_{\rm tot}$ is not linear (see Fig.\,\ref{pic:model}). 
Thus, the procedure of neutrino luminosity estimation is divided into a few steps:
(i) using the apparent photon luminosity of a NS we get the photon luminosity at the NS surface
using~(\ref{eq:red_shift}),
(ii) the photon luminosity at the NS surface is converted into neutrino luminosity at the NS using approximation (\ref{eq:approx}) or more accurate estimations \citep{2018MNRAS.476.2867M},
(iii) we get neutrino luminosity at the infinity accounting for the gravitational redshift~(\ref{eq:red_shift}). 
Because the apparent luminosity can be different from the actual one due to the geometrical beaming \citep{2017MNRAS.468L..59K,2021MNRAS.501.2424M}, we illustrate the influence of this uncertainty assuming that the difference between two photon luminosities is within a factor of 2.
With the known neutrino luminosity corrected for the gravitational redshift and distance to the source, we can estimate the expected neutrino energy flux at Earth.

\section{Estimations of neutrino luminosity}
\label{sec:Results}

Estimating neutrino luminosity, we assume NS mass and radius to be $M=1.4M_\odot$ and $R=10^6\,{\rm cm}$ respectively.
All ULX pulsars are known to be strongly variable (see discussion on the observed variability
in M82~X-2 by \citealt{2016MNRAS.457.1101T},
in NGC~5907~X-1 by \citealt{2017Sci...355..817I}, 
in NGC~1313~X-2, M51~X-7, NGC~7793~P13 and NGC~300~X-1 by \citealt{2021A&A...649A.104G}) and our estimations of their neutrino luminosity and corresponding neutrino energy flux are based on the maximal detected X-ray luminosity.

The estimated neutrino luminosity and neutrino energy flux $F_\nu$ in six pulsating ULXs and in three bright Be X-ray transients at the peak of their outbursts are represented in Table\,\ref{tab:ULXs}.
Despite relatively large neutrino luminosity expected in the ULXs, due to the difference in distance to the sources, the neutrino energy flux at Earth is below $10^{-5}\,{\rm MeV\,cm^{-2}\,s^{-1}}$, which is smaller than the neutrino flux expected from considered Be X-ray transients. 

\begin{figure}
\centering 
\includegraphics[width=8.7cm]{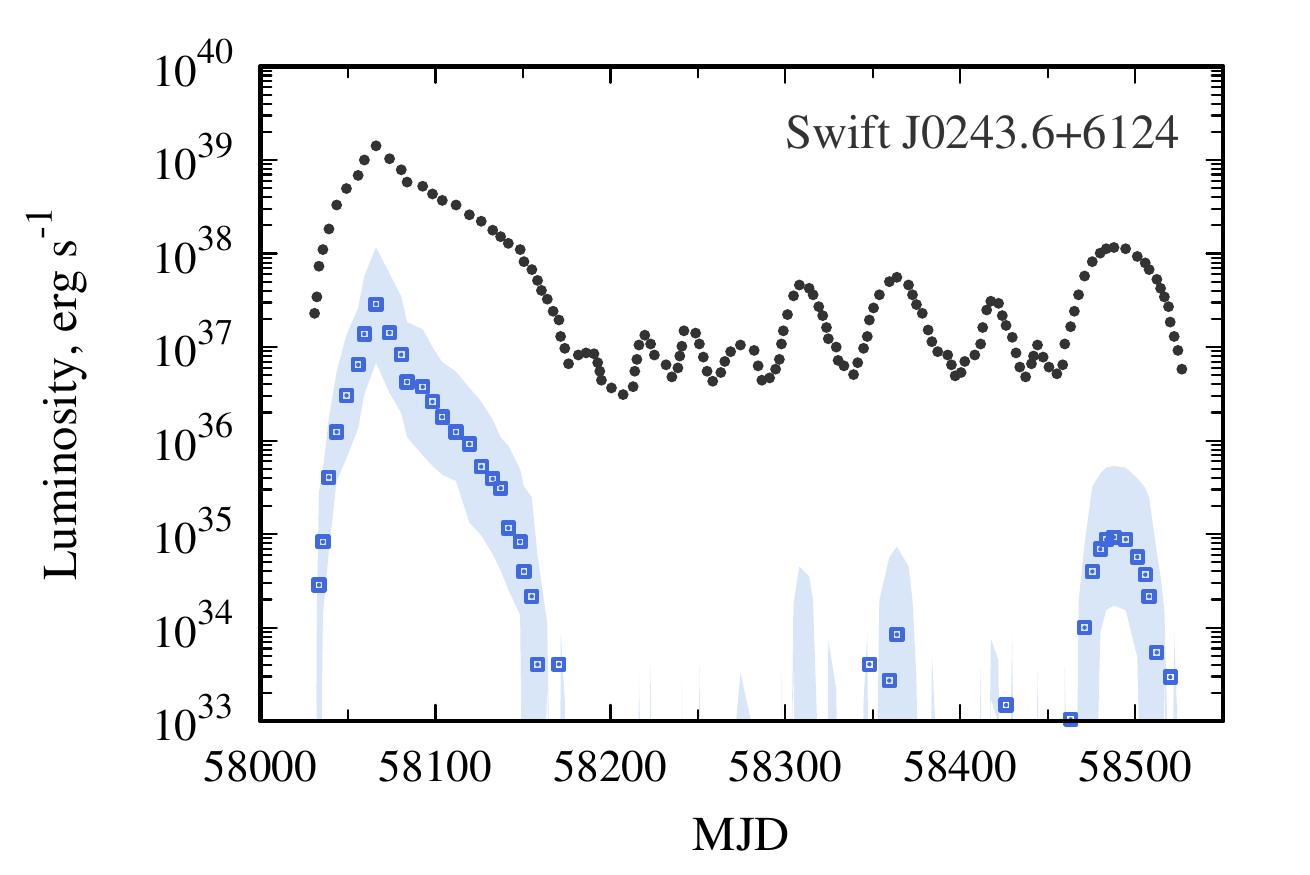} 
\caption{
Black circles represent the photon light curve of \swift during its outburst in 2017-2018 reproduced on the base of data from \citealt{2020MNRAS.491.1857D}, while blue squares represent neutrino light curve calculated on the base of \citealt{2018MNRAS.476.2867M}.
}
\label{pic:sc_swift}
\end{figure}

\begin{figure}
\centering 
\includegraphics[width=8.7cm]{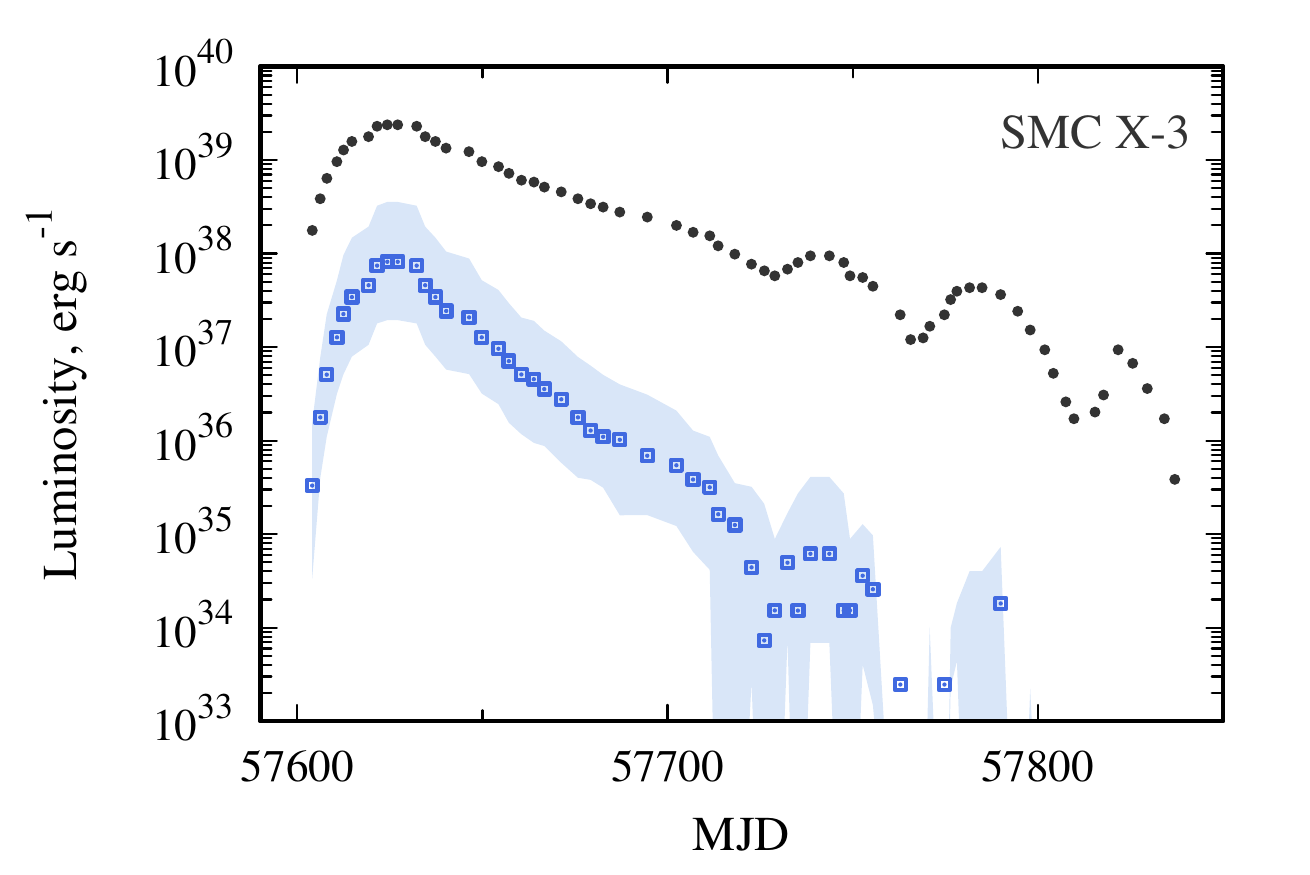} 
\caption{
Black circles represent the photon light curve of SMC~X-3 during its outburst in 2016-2017 reproduced on the base of data from \citealt{2017A&A...605A..39T}, while blue squares represent neutrino light curve calculated on the base of \citealt{2018MNRAS.476.2867M}.
}
\label{pic:sc_smc}
\end{figure}

\begin{figure}
\centering 
\includegraphics[width=8.7cm]{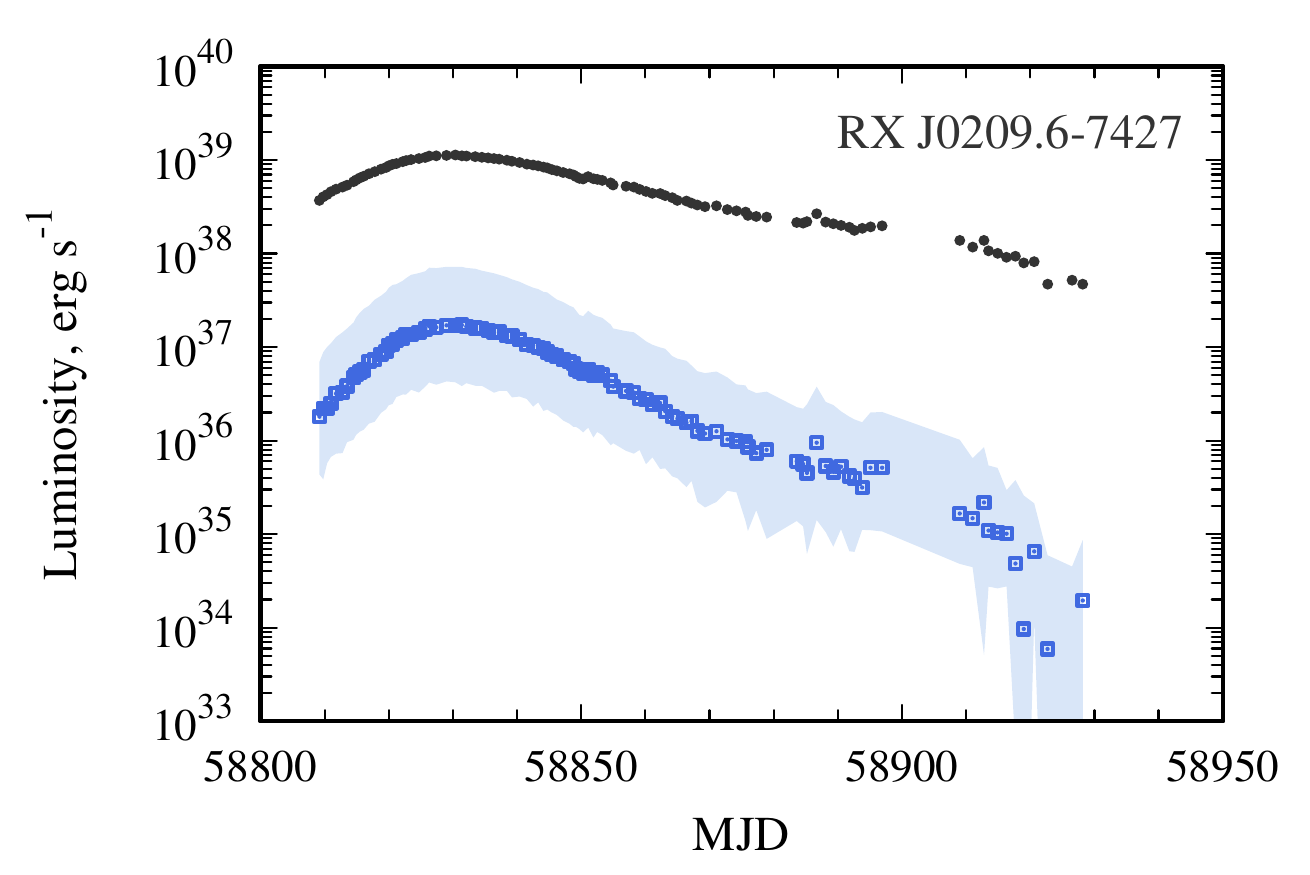} 
\caption{
Black circles represent the photon light curve of \RX during its outburst in 2019-2020 reproduced on the base of data from \citealt{2022ApJ...938..149H} , while blue squares represent neutrino light curve calculated on the base of \citealt{2018MNRAS.476.2867M}.
}
\label{pic:sc_rx}
\end{figure}

By using recent observations of three bright X-ray transients - \swift, SMC~X-3 and \RX- during their outbursts, we have obtained neutrino light curves during the outbursts (see blue squares in  Fig.\,\ref{pic:sc_swift},\,\ref{pic:sc_smc} and \ref{pic:sc_rx}). 
Because the neutrino luminosity is strongly dependent on the total luminosity of accreting NS, the uncertainties in estimations of actual X-ray luminosity result in significant uncertainties in estimations of neutrino luminosity and energy flux.
Blue regions in Fig.\,\ref{pic:sc_swift},\,\ref{pic:sc_smc},\,\ref{pic:sc_rx} illustrate the uncertainty in neutrino luminosity due to the uncertainty in photons luminosity within a factor of 2. 
The majority of neutrinos are emitted during the main peak of the outbursts in Be XRPs. 
When integrating the neutrino light curve obtained for three Be transients, we estimated the total energy emitted with neutrinos during the outbursts: 
$\sim5\times 10^{43}\,{\rm erg}$ in \swift,
$\sim1.7\times 10^{44}\,{\rm erg}$ in SMC~X-3,
and
$\sim 4.5\times 10^{43}\,{\rm erg}$ in \RX.

The estimations of neutrino luminosity are dependent on NS mass and radius because of different contributions of the gravitational redshift.
However, the uncertainties due to the unknown mass and radius of a NS are much smaller than the uncertainties due to the relation between the actual and apparent luminosity of the objects. 

\section{Neutrino flux property and detection}
\subsection{Neutrino composition and {energy spectrum}}

{
The possibility of neutrino detection at Earth is determined not only by the neutrino flux $F_\nu$ but also by neutrino species composition and their spectral distribution.} 
The dominant process of neutrino emission from relatively hot electron-positron plasma is annihilation of electron-positron pairs (see, e. g.,~\citealt{1992PhRvD..46.3256K}). 
{The result of this reaction is production of neutrino and antineutrino of one of three flavors: }
electron ($\nu_e$, $\overline\nu_e$), 
muon ($\nu_\mu$, $\overline\nu_\mu$), 
or tau ($\nu_\tau$, $\overline\nu_\tau$).
{In the case of non-relativistic plasma and under the condition of extremely strong magnetic field $B \gg B_0 \approx 4.41 \times 10^{13}$~G, neutrino emissivity {due to the} annihilation is proportional to $c_V^2 + c_A^2$, while in the case of $B \ll B_0$,
the emissivity of proportional to $c_V^2$ \citep{1992PhRvD..46.4133K}.}
Here $c_V = \pm 1/2 + 2 \sin^2 \theta_W$ and $c_A = \pm 1/2$
are {the vector and axial constants of the charged lepton current,} 
and {$\sin^2 \theta_W \approx 0.2312$, where $\theta_W$ is the Weinberg angle} \citep{Workman:2022ynf}.
The upper (plus) signs in $c_V$ and $c_A$ correspond to the electron flavor, while the lower (minus) signs correspond to muon and tau species.
So we could reconstruct the {initial} contributions of a different flavor to summarize neutrino luminosity {$L_\nu$.
The initial electron neutrino and antineutrino luminosities are}
\beq
L_{\nu_e} \!\! = L_{\overline\nu_e} \!\! \equiv f_{0e} \, L_\nu
\eeq
and {the initial muon/tau neutrino and antineutrino luminosities are}
\beq
L_{\nu_\mu} \!\! = L_{\overline\nu_\mu} \!\! = L_{\nu_\tau} \!\! = L_{\overline\nu_\tau} \!\! 
\equiv f_{0x} \, L_\nu,
\eeq
where $f_{0e} \approx 0.35$, $f_{0x} \approx 0.075$ for {the case of magnetar-lile magnetic field strength} $B \gg B_0$,
and $f_{0e} \approx 0.4985$, $f_{0x} \approx 0.00075$ for {the case of relatively weak magnetic fields} $B \ll B_0$ \citep{1992PhRvD..46.4133K}. 
We neglect neutrino oscillation in the matter of X-ray source due to the essential uncertainty of accretion column geometry and its small extent.
{Gravitational redshift does not affect the flavor composition of neutrino flux.}

{However, neutrino propagation over the distance between the source and observer affects the composition of the flux due to the vacuum neutrino oscillations.}
The initial and final flavor fractions are related as~\citep{Mena:2006eq}:
\begin{eqnarray}
\begin{gathered}
f_e = f_{0e} - (f_{0e} -  f_{0x}) \, \sin^2(2\theta_{12}) / 2 ,
\\ 
f_x = f_{0x} - (f_{0e} -  f_{0x}) \, \sin^2(2\theta_{12}) / 4 ,
\end{gathered}
\end{eqnarray}
where $\sin^2\theta_{12} \approx 0.307$~\citep{Workman:2022ynf} is a squared sin of the neutrino mixing angle.
Hence we obtain 
$f_e \approx 0.23$, $f_x \approx 0.135 $
for magnetar-like accretor and
$f_e \approx 0.29$, $f_x \approx 0.105$ for {the accretor of weaker surface magnetic field}.
These values are equal approximately, {so we used the following averages:}
\beq
\label{eq:fractions}
f_e = 1/4 , 
\quad
f_x = 1/8 .
\eeq

To restore the neutrino spectrum, we follow the result of paper by \citealt{Misiaszek:2005ax}.
Possibility of neutrino detection could be obtained from {the specific neutrino flux $\varphi_\nu$, which has the dimension $[\varphi_\nu]=\rm{[cm^{-2} s^{-1} MeV^{-1}]}$.
The specific flux} is related to neutrino energy flux $F_\nu$ as
\beq
F_\nu = \int\limits_0^\infty \varepsilon \, \varphi_\nu(\varepsilon) \, d \varepsilon ,
\eeq
where $\varepsilon$ is neutrino energy measured in MeV.
Following the result of~\citealt{Misiaszek:2005ax}, we approximate neutrino flux at Earth from an X-ray source as
\beq
\varphi_\nu(\varepsilon) = 
\frac{s^s F_\nu}{\varGamma(s) \; \overline\varepsilon^2}
\left(\frac{\varepsilon}{\overline\varepsilon} \right)^{s-1}
\exp \left( - s \frac{\varepsilon}{\overline\varepsilon} \right) ,
\eeq
where the parameter $s \approx 4.181$ and the mean neutrino energy $\overline\varepsilon = m_e + 3/2 \, T_X$, 
$T_X$ is the temperature of neutrino emitted medium and
$m_{\rm e} \approx 0.511$~MeV is electron mass.
Note, that the spectra of different neutrino flavors are the same due to the kinematics of the annihilation process, and functions $\varphi_{\nu_i}$ corresponding to different flavors are differed from $\varphi_\nu$ 
by a factor $f_e$ or $f_x$ only~(\ref{eq:fractions}).
\begin{figure}
\centering 
\includegraphics[width=8.cm]{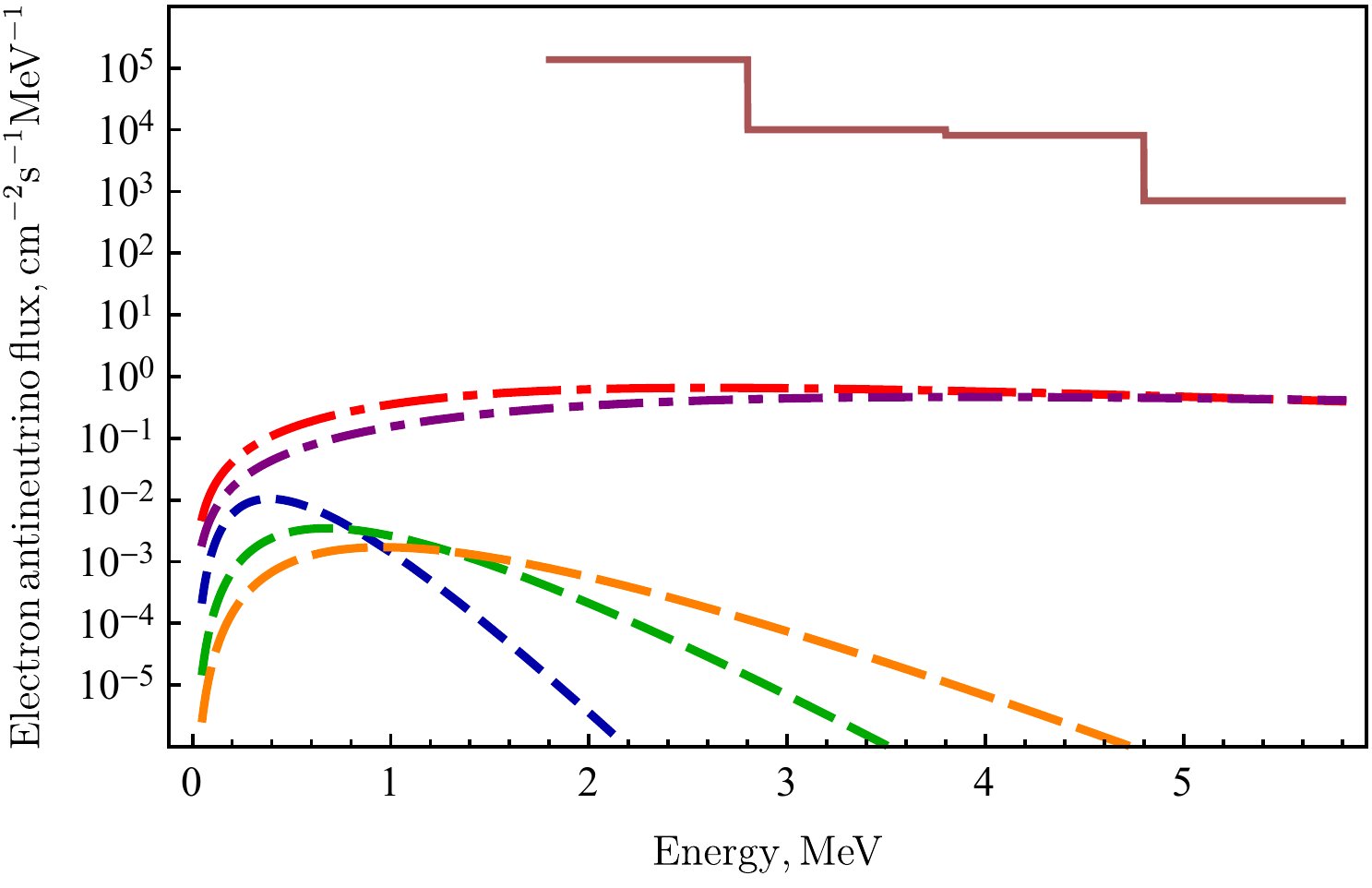} 
\caption{
Set electron antineutrino fluxes $\varphi_{\overline\nu_e}$ as a function of energy.
Swift J0243.6+6124: curves for different temperatures of neutrino emitted area
$T_X \!=\! 0$~MeV (blue short dashed), 
$T_X \!=\! 0.25$~MeV (green dashed),
$T_X \!=\! 0.5$~MeV (orange long dashed)
Diffusion supernova background:
pure neutron star $T_{\rm SN} \!=\! 4$~MeV (red dot and long dash) 
and pure black hole $T_{\rm SN} \!=\! 6$~MeV (purple dot and dash) forming.
Borexino model-independent limit on electron antineutrino fluxes from unknown sources (brown solid).
}
\label{pic:Union}
\end{figure}
 
The dependence of the neutrino spectrum on temperature is shown in Fig.~\ref{pic:Union}.
{One can see that} the spectrum of neutrino emitted {by hotter material} is wider but lower at {its} maximum value.

\begin{figure*}
\centering 
\includegraphics[width=14.5cm]{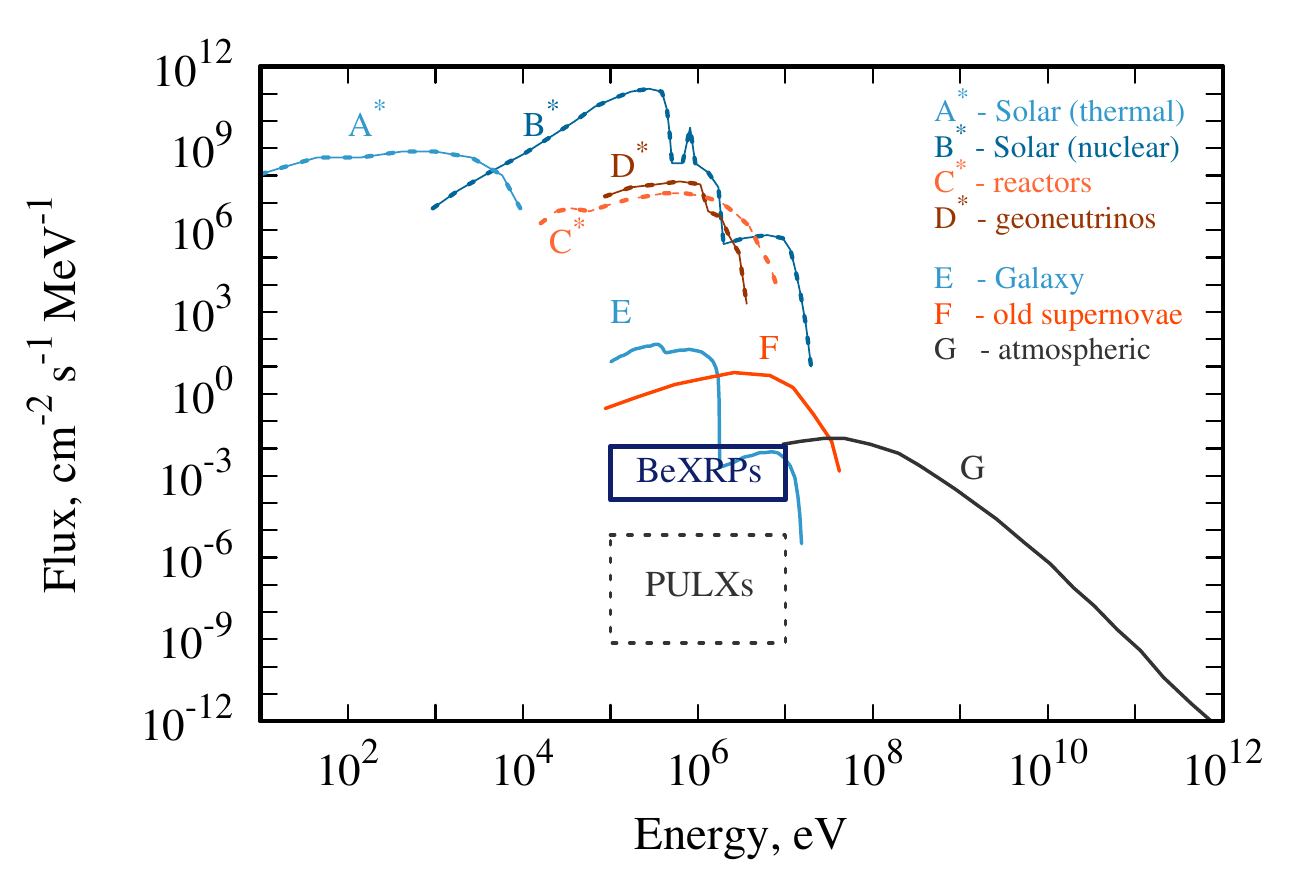} 
\caption{
{
The energy spectrum of neutrino background at Earth summed over flavors in the energy range from $10\,{\rm eV}$ to $10^{12}\,{\rm eV}$ (taken from \citealt{Vitagliano:2019yzm} and \citealt{1998A&A...333..910B}).
The background is composed of 
solar thermal (A$^*$) and nuclear neutrinos (B$^*$), 
neutrinos from nuclear reactors (C$^*$), 
geoneutrinos (D$^*$), 
isotropic component of neutrino flux from stars of the Galaxy (E),
neutrinos from old supernovae (F), 
and atmospheric neutrinos (G).
Dashed lines represent non-isotropic neutrino background, while the background represented by solid lines is expected to be isotropic. 
Note, that the total neutrino flux due to the stars of the Galaxy is not isotropic and expected to be larger (by a factor $\sim 8$ in comparison to the isotropic component) in the direction towards the Galactic center (see Fig.\,8 in \citealt{1998A&A...333..910B}).   
Two rectangles show the area, where we expect neutrino flux from three bright BeXRPs at the peak of their outbursts (solid blue) and from six pulsating ULXs (dashed black).
}
}
\label{pic:sc_neutrino_sp}
\end{figure*}

\subsection{Neutrinos detection}

As seen from Table~\ref{tab:ULXs}, pulsating ULXs have {higher} neutrino luminosity {in comparison to three bright Be XRPs considered in the paper}.
However, {the neutrino energy flux at the Earth is smaller} due to the large distance to ULXs.
Below we discuss the possibility of neutrino detection  
\swift, which neutrino energy flux at the Earth is expected to the the highest one.

{
The observability of accreting NSs in neutrinos is determined by the sensitivity of neutrino telescopes and the ratio of neutrino flux from sources to the background. 
The sources of neutrino background can be isotropic and non-isotropic.
Non-isotropic neutrino background in the energy band from $0.01$ to $100\,{\rm MeV}$ is dominated by solar thermal and nuclear neutrinos (Fig.\,\ref{pic:sc_neutrino_sp}, A$^*$ and B$^*$ components), 
neutrinos from the ground based reactors (Fig.\,\ref{pic:sc_neutrino_sp}, C$^*$)
and geoneutrinos (Fig.\,\ref{pic:sc_neutrino_sp}, D$^*$).
The isotropic neutrino flux in the same energy band is due to neutrino emission of the nearest stars of the Galaxy (Fig.\,\ref{pic:sc_neutrino_sp}, E), 
old core-collapse supernova explosions  (Fig.\,\ref{pic:sc_neutrino_sp}, F)
and atmospheric neutrinos (Fig.\,\ref{pic:sc_neutrino_sp}, G).
}

{
Non-isotropic neutrino background is dominated over the isotropic background at a MeV energy range.
Solar nuclear neutrino flux} is estimated to be 
$$F_\odot \gtrsim 10^8\,{\rm MeV\,cm^{-2}\,s^{-1}}.$$ 
{
Decay of radioactive isotopes in the Earth provides the flux of terrestrial antineutrinos \citep{2012RvGeo..50.3007D}
$$F_\oplus\sim 10^6\,{\rm MeV\,cm^{-2}\,s^{-1}}.$$ 
Nuclear reactors release a few percents of their total energy production in the form of MeV-range antineutrinos and cause the flux \citep{2011PhRvC..83e4615M}
$$F_{\rm nucl}\sim F_\oplus\sim 10^6\,{\rm MeV\,cm^{-2}\,s^{-1}}.$$}

{
Isotropic neutrino background (its components are shown by solid lines in Fig.\,\ref{pic:sc_neutrino_sp}) is a few orders of magnitude lower in comparison to the non-isotropic one.}
Ancient core collapse supernovae provide the neutrino flux 
$$F_{\rm SN}\sim 1\,{\rm MeV\,cm^{-2}\,s^{-1}}.$$ 
{
Neutrino flux from stars of the Galaxy is strongly dependent on energy (see curve E in Fig.\,\ref{pic:sc_neutrino_sp}) and direction: the isotropic component due to the neutrino emission by the nearest stars is estimated to be almost an order of magnitude smaller than the flux in the direction towards the Galactic center \citep{1998A&A...333..910B}.
The composition of Galactic neutrinos is similar to the composition of neutrino flux from the Sun.
}
The atmospheric neutrino {and antineutrino flux due to interaction of cosmic rays with the atmosphere of Earth is estimated to be}
$$F_{\rm atm}< 10^{-2}\,{\rm MeV\,cm^{-2}\,s^{-1}}$$
at a MeV energy range \citep{2015PhRvD..92b3004H}.

Detecting electron neutrinos from an X-ray source is difficult due to significant background from {the Sun and Galactic stars}.
Thus, observation of accreting NSs is more preferred in the other species of neutrino and antineutrino {($\nu_\mu\,/\overline{\nu}_\mu$ or $\nu_\tau/\overline{\nu}_\tau$)}.
{The most significant isotropic neutrino background of mixed flavours} is provided by old core-collapse supernovae {explosions}, which are sources of all neutrino species.
Although this background has an average energy of $\sim\!\! 6$~MeV, its estimated neutrino flux~\citep{Vitagliano:2019yzm}:
\beq   
\begin{aligned}
\varphi_{\rm SN}(\varepsilon) \approx \, & 5.12  
\left(\frac{4\,\text{MeV}} {T_{\rm SN}}\right)^2 
\arctan\left[3 \left(\frac{\varepsilon}{T_{\rm SN}}\right)^{3/2}\right]
\\ 
& \times e^{-{1.03 \varepsilon}/{T_{\rm SN}}}
\end{aligned}
\eeq 
exceed value corresponding to Swift J0243.6+6124.
{Supernova} neutrino flavor composition {is expected to be different for the explosions leading to production of NSs and black holes \citep{2006PhRvL..97i1101S,2020PhRvD.101l3013W}.
In the case of NS remnant, the flavour composition is given by} 
$f_e \!\approx\! 1/4$, $f_x \!\approx\! 1/8$, and 
$T_{\rm SN}\approx 4\,{\rm MeV}$, 
{while in the case of black hole production neutrinos are dominated by the electron flavour:}
$f_e \!\approx\! 2/5$, $f_x \!\approx\! 1/20$, and  
$T_{\rm SN}\approx 6\,{\rm MeV}$ \citep{Vitagliano:2019yzm}.
As one can see from the Fig.~\ref{pic:Union}, the diffusion supernova background is in excess of the electron antineutrino emission from Swift J0243.6+6124 even in a MeV energy band. 
A similar result is for other neutrino species. 
Thus, old core-collapse supernova could produce significant background even for the brightest in neutrino X-ray source Swift J0243.6+6124.

{Detection of MeV neutrinos and antineutrinos requires use of underground detectors.}
Further, we estimate the possibility of neutrino detection from X-ray sources using data from the Borexino underground detector \citep{2009NIMPA.600..568A}.
We use a model-independent upper limit in the energy range $1.8-16.8\,{\rm MeV}$ on electron antineutrino fluxes from unknown sources~\cite{Borexino:2019wln}.
In this paper, the monochromatic neutrinos model with 1 MeV wide bins was used, so the upper limit is a piecewise function. 
The result is shown in Fig.~\ref{pic:Union}.
As one can see from the picture, the Borexino model-independent limit exceeds electron antineutrino flux from Swift J0243.6+6124 by more than eight orders of magnitude.
Note that the Borexino limit obtains through detector neutrino events of inverse beta decay of the free proton.
The energy threshold of this reaction is 1.8~MeV, so the maximum neutrino spectrum from X-ray sources occurs in the lower energy range.
Therefore the reaction of neutrino electron scattering is preferable for the detection of neutrinos from an X-ray source. 
Moreover, all neutrino species are involved in this reaction, in contrast to the inverse beta decay, where only the electron antineutrino participates.
All backgrounds except old supernovae are electron neutrinos or antineutrinos. 
Thus, muon and tau (anti)neutrinos detection is the preferable strategy for the detection of neutrino signal from bright XRPs.

As discussed above, even neutrino brightest X-ray pulsar \swift has the flux
(i) under the background of diffuse neutrino from old supernovae, and
(ii) significantly less than the reliable threshold of neutrino detection by existent detectors.
Finally, we estimate the possibility of neutrino detection from near the Earth but unknown at present X-ray sources.
For this purpose, we obtain the following simple approximation data from Table~\ref{tab:ULXs}:
$
L_\nu \approx 1.6 \times 10^{39} ( {L_X}/{10^{40}\,\text{erg s}^{-1}} )^2 .
$
Note that approximation values differ from data less than 50~\%.
Hence we obtain an estimation of the distance on which neutrinos from the X-ray sources could be detected at the maximum of their spectrum:
\beq
\label{eq:Ddetect}
D_* \sim 
\left( \frac{10^5 \, \text{cm$^{-2}$s$^{-1}$MeV$^{-1}$}}{\varphi_{th}} \right)^{1/2}
\left( \frac{L_{\rm X}}{10^{39}\,\ergs} \right)\,{\rm ly},
\eeq
where $\varphi_{th}$ is threshold of detection neutrino flux.
It follows from the estimation that Be XRPs, able to be detected by existent neutrino telescope, should be located among the nearest stars. 
Whereas pulsating ULXs could be detected from a distance of a few hundred light years.
Note that a special method of selecting neutrino events in the detector, such as taking into account the direction of X-ray sources, detection muon and tau (anti)neutrinos, could decrease the detection threshold and increase the distance to the detectable X-ray objects.

\section{Summary}
\label{sec:Summary}

In this paper, we have considered accreting strongly magnetized NSs as possible sources of neutrino emission.
Applying the model proposed by \citealt{2018MNRAS.476.2867M} to six ULX pulsars and three bright Be XRPs, we have estimated neutrino luminosity and neutrino energy flux from these objects (see Table\,\ref{tab:ULXs}).
{Despite the large total neutrino luminosity of ULX pulsars, the neutrino energy flux from the Be X-ray transients of our Galaxy (\swift) and Magellanic Clouds (SMC~X-3 and \RX) is expected to be dominant due to a relatively small distance from them.}
For the case of Be XRPs, we have provided the expected neutrino light curve during their recent outbursts (see Fig.\,\ref{pic:sc_swift},\,\ref{pic:sc_smc} and \ref{pic:sc_rx}). 
The total energy release due to the neutrino emission during the outbursts was estimated to be at the level of $10^{44}\,{\rm erg}$.

{Because the major process of neutrino emission in bright XRPs is the annihilation of electron-positron pairs, the flux is composed equally of neutrinos and antineutrinos of a MeV energy range, where about a half of particles belong to the electron flavour.
In a MeV energy range, the neutrino background is dominated by non-isotropic sources: solar neutrinos, antineutrinos from nuclear reactors and terrestrial antineutrinos due to the decay of radioactive isotopes (B$^*$,C$^*$,D$^*$ curves in Fig.\,\ref{pic:sc_neutrino_sp} respectively).
Isotropic neutrino background is a few orders of magnitude lower and in a MeV energy range dominated by neutrinos/antineutrinos from old core-collapse supernova explosions and neutrinos from the nearest stars of the Galaxy (F and E curves in Fig.\,\ref{pic:sc_neutrino_sp} respectively).
Because there is a lack of antineutrinos in the flux from the Sun and stars of the Galaxy, it is preferable to observe accreting NSs in antineutrinos.}
However, the flux from bright Be X-ray transients considered in our paper is still more than two orders of magnitude below the isotropic neutrino background due to old supernova explosions (see Fig.\,\ref{pic:sc_neutrino_sp}). 
{
Therefore, direct registration of neutrino emission from accreting NSs seems impossible at present and confirmation of significant energy losses with neutrino by NSs requires the development of indirect methods.
}

\section*{Acknowledgements}

AAM thanks UKRI Stephen Hawking fellowship.
We are grateful to an anonymous referee for useful comments and suggestions. 

\section*{Data availability}

The calculations presented in this paper were performed using a private code developed and owned by the corresponding author. All the data appearing in the figures are available upon request. 


{

}

\bsp 
\label{lastpage}
\end{document}